\documentclass[10pt]{kluwer}
\usepackage{epsfig}
\begin{document}
\begin{article}
\begin{opening}
\title{Evidence and Implications of Pressure Fluctuations in
the ISM}
\author{Edward B. \surname{Jenkins}\email{ebj@astro.princeton.edu}}
\institute{Princeton University Observatory, Princeton, NJ 08544-1001,
USA}
\runningtitle{Pressure Fluctuations}
\runningauthor{Jenkins}
\begin{abstract}
A recent survey of the fine-structure excitation of neutral carbon
reveals that the interstellar medium in the Galactic plane exhibits a
thermal pressure, $nT/k$, that ranges from about $10^3$ to $10^4\,{\rm
cm}^{-3}\,$K from one location to the next, with occasional excursions
in excess of about $10^5\,{\rm cm}^{-3}\,$K.  The large excitations for
small amounts of gas indicate that some regions are either subjected to
shocks or must be pressurized within time scales much shorter than the
time needed to reach thermal equilibrium.  These rapid fluctuations
probably arise from the cascade of macroscopic mechanical energy to
small scales through a turbulent cascade.  One consequence of this
effect is that changes in gas temperature can arise from near adiabatic
compressions and expansions, and this may explain why investigations of
21-cm emission and absorption reveal the presence of hydrogen at
temperatures well below the expected values derived from the balance of
various known heating and cooling processes. 
\end{abstract}
\abbreviations{\abbrev{ISM}{Interstellar Medium}}
\end{opening}

\section{Introduction}\label{intro}

While gases within the Milky Way show enormous ranges in density and
temperature, the simplest, most general picture is that variations in
the product of these two quantities, the thermal pressure, are very much
smaller by comparison.  This may be true at the most superficial level,
and indeed a convincing theory (at the time) of the ISM  (Field,
Goldsmith \& Habing 1969) was built upon the premise that the observable
phases of the ISM arise simply from a thermal instability  (Field 1965)
with uniform heating at a nearly constant pressure.  In this picture,
the cool phase is reasonably static and confined by a uniform external
pressure from a surrounding warm medium.  It is now more apparent that
this tranquil picture of the ISM is probably unjustified since we have
known for some time that disturbances over macroscopic scales arising
from supernova explosions, newly formed H~II regions, stellar mass loss,
bipolar flows, infalling gas from the halo and Galactic density wave
shocks can all inject significant mechanical energy in various locations
and disrupt the medium.

A major refinement in the theory of the ISM was created by McKee \&
Ostriker  (1977), who recognized some important global consequences of
energetic phenomena, which can lead to the creation of large filling
factors for material at high temperatures and low densities [see also 
(Cox \& Smith 1974; Cox 1979, 1981)].  An often overlooked property of
the McKee \& Ostriker model [e.g.,  (Cox 1995)] is that the pressures
should vary by significant amounts from one location to the next 
(Jenkins, Jura \& Loewenstein 1983).  Ultimately, we should expect that
large-scale enhancements and rarefactions in the pressures should become
more widespread and chaotic as they propagate to different locations and
migrate to smaller scales through the creation of a turbulent cascade 
(Mac Low et al. 2001).

Thermal pressures are a minor component of the total ISM pressure, which
also includes magnetic, dynamical (i.e., turbulent) and cosmic ray
components.  Boulares \& Cox  (1990) estimated that the weight of
material on either side of the Galactic plane should balanced by a total
pressure $p/k=2.5\times 10^4\,{\rm cm}^{-3}\,$K.  Inasmuch as typical
thermal pressures are about a factor 10 lower and thus seemingly of
little importance dynamically, one might wonder why we are interested in
their average value and distribution over different volumes of space. 
The answer arises from several considerations: first, thermal pressure
values are molded by processes arising from the other forms of pressure
and thus convey important information, second, they reflect the physical
state of the gas at the atomic scale and thus have an influence on
various reaction rates and equilibria, such as those pertaining to
ionization, heating, cooling and chemistry.  Finally, thermal pressures
are easy to measure.

There are various ways to determine thermal pressures, but probably the
most straightforward one is by measuring the relative populations of
atoms in the excited fine-structure levels of their ground electronic
states.  This can be done by observing absorption lines of the atomic
multiplets, seen in the uv spectra of hot background stars.  The fine-structure levels are populated by collisions with other atoms, ions,
molecules, and electrons.  The excitations (and de-excitations) occur at
rates proportional to the local densities of these collision partners,
and a balance between these processes and the spontaneous radiative
decay of the upper levels establishes an equilibrium (with a very short
time constant) determined by the local density and temperature of the
gas.

\section{Thermal Pressures from C~I Fine-structure Excitation}

Of the various species that have ground electronic states with two or
more fine-structure levels, in most circumstances neutral carbon seems
to be the most useful one for indicating local conditions.  While one is
not able to determine unique values for either temperature or density,
Jenkins \& Shaya  (1979) showed that under the most common conditions in
the ISM their product was reasonably well defined from the two
population ratios of the upper levels with respect to the total in all
three states.  Moreover, since the problem is over-determined by having
more than one ratio, it is possible to differentiate between a uniform,
medium pressure and an admixture of two or more regions having low and
high pressures, even if they have the same radial velocity.  The latter
property was exploited by Jenkins \& Tripp  (2001) (hereafter JT01) to
show that there was pervasive evidence for a small amount of gas at $p/k
> 10^5\,{\rm cm}^{-3}\,$K, some two orders of magnitude above the
general average pressure.

\begin{figure}
\centerline{\psfig{file=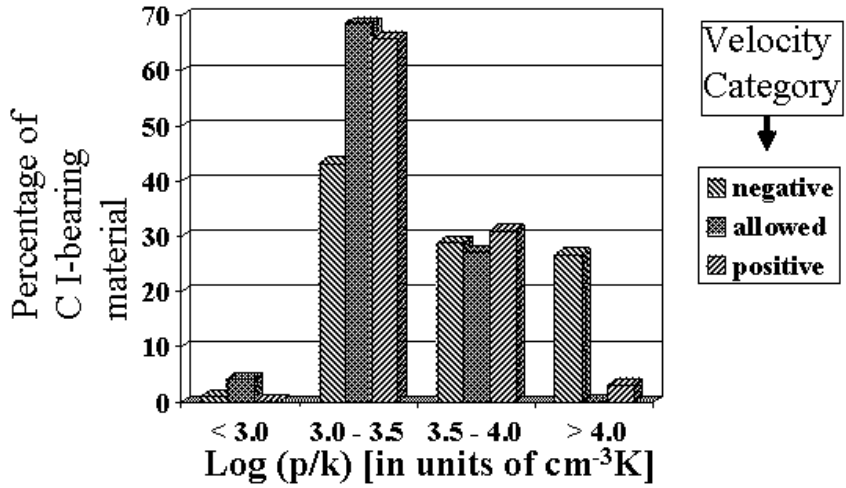,width=15cm,angle=0}}
\caption{A histogram indicating the relative amounts of C~I in different
logarithmic intervals found by Jenkins \& Tripp  (2001), with the gas
segregated according to whether the radial velocity is below, within, or
above the allowed range of velocities between the observer and the star,
as determined from estimates for the effects of differential Galactic
rotation along the respective sight lines.}\label{dist_pressures}
\end{figure}

Figure~\ref{dist_pressures} shows the amounts of C~I within different
pressure intervals that were found in a survey of 21 stars by JT01, who
used the highest resolution mode (E140H with the narrowest entrance
slit, yielding $R$ = 200,000) of the {\it Space Telescope Imaging
Spectrograph\/} on {\it HST}.  The pressure distributions show a clear
link to the kinematical properties of the gas: disturbed material
outside the range of normally allowed radial velocities are dominant at
high pressures, while the lowest extremes in pressure appear to be
quiescent.  From this it seems clear that the thermal pressures are
responding to dynamical processes, as we would expect.  The
preponderance of negative velocity gas at high pressures probably arises
from material compressed by mass flows from the target stars or their
neighbors in the early-type stellar associations.

\begin{figure}
\centerline{\psfig{file=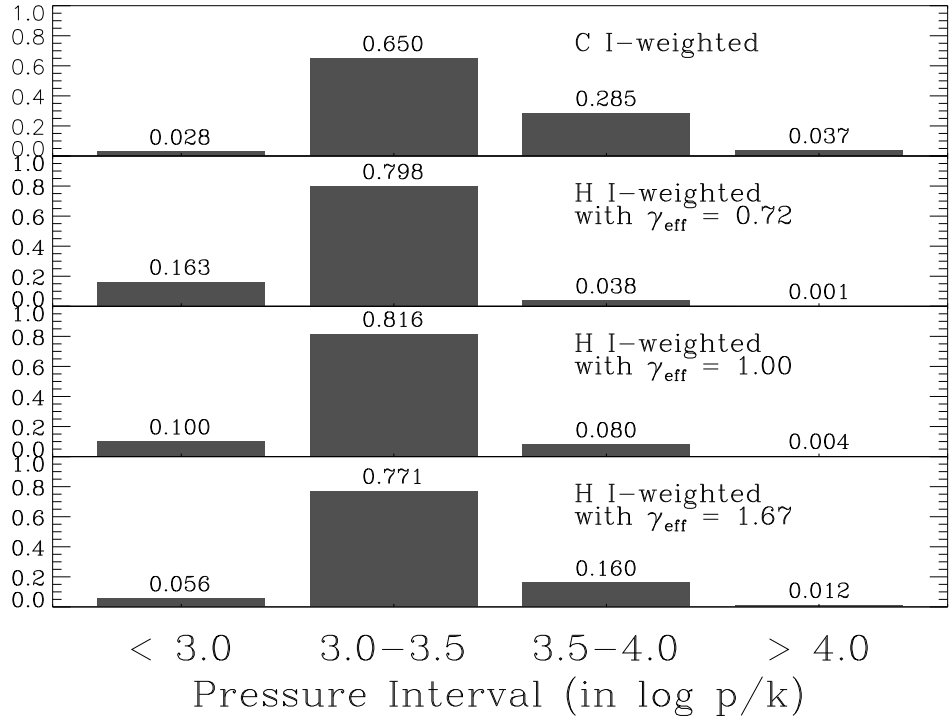,width=14cm,angle=0}}
\caption{Transformations of the original C~I pressure distribution (top
panel: sum of all velocity intervals depicted in
Fig.~\protect\ref{dist_pressures}) to the total amounts of gas for 3
values of the barytropic index $\gamma_{\rm eff}$ (lower 3 panels),
derived by applying Eq.~\protect\ref{HI_corr} to each pressure interval. 
The three values of $\gamma_{\rm eff}$ represent (1) 0.72: approximate
slope of the thermal equilibrium relation, 1.00: isothermal gas, 1.67
purely adiabatic behavior.}\label{corr_dist_pressures}
\end{figure}

As we consider the evidence from C~I, we must acknowledge that the
distribution of pressures reflects material weighted by the respective
abundances of C~I, which in turn are governed by the ionization
equilibrium of C~I with the more common form of carbon, C~II.  This
effect exaggerates the amount of high-pressure material because there is
a shift toward greater C~I/C~II at high densities.  If the gas can be
characterized with an equation of state (see \S\ref{effects}) $p\propto
n^{\gamma_{\rm eff}}$, where $\gamma_{\rm eff}$ is the barytropic index,
\begin{equation}\label{HI_corr}
n({\rm H~I})\propto p^{0.324-1.622/\gamma_{\rm eff}} n({\rm C~I})~.
\end{equation}
Eq.~\ref{HI_corr} shows the correction\footnote{The coefficients in the
exponent of $p$ are derived from the calculations of carbon ionization
equilibria by Jenkins (2002) which include not only recombinations of
ions with free electrons but also their neutralization by very small,
negatively-charged dust grains  (Weingartner \& Draine 2001).} that must
be applied to the C~I pressure distribution function to obtain an
even-handed representation for all of the neutral gas.  The
transformations of the distributions for three different values of
$\gamma_{\rm eff}$ are shown in Fig.~\ref{corr_dist_pressures}.

\section{The Effects of Changes in Pressure}\label{effects}

We now imagine what happens when the ISM is in a chaotic state where
dynamical processes force the pressures to vary with position and time
on either side of some mean value.  In this picture, we may characterize
the pressure fluctuations as random.  Even though temperatures and
densities may vary, we must still insist that a balance between heating
and cooling is maintained over a time interval much longer than the
characteristic thermal time scales.  Of course, our assumption about
randomness is an idealization: in large part the real data will reflect
this condition, but from time to time coherent effects (pressurizations)
from specific sources are bound to manifest themselves and distort the
results.

\begin{figure}
\centerline{\psfig{file=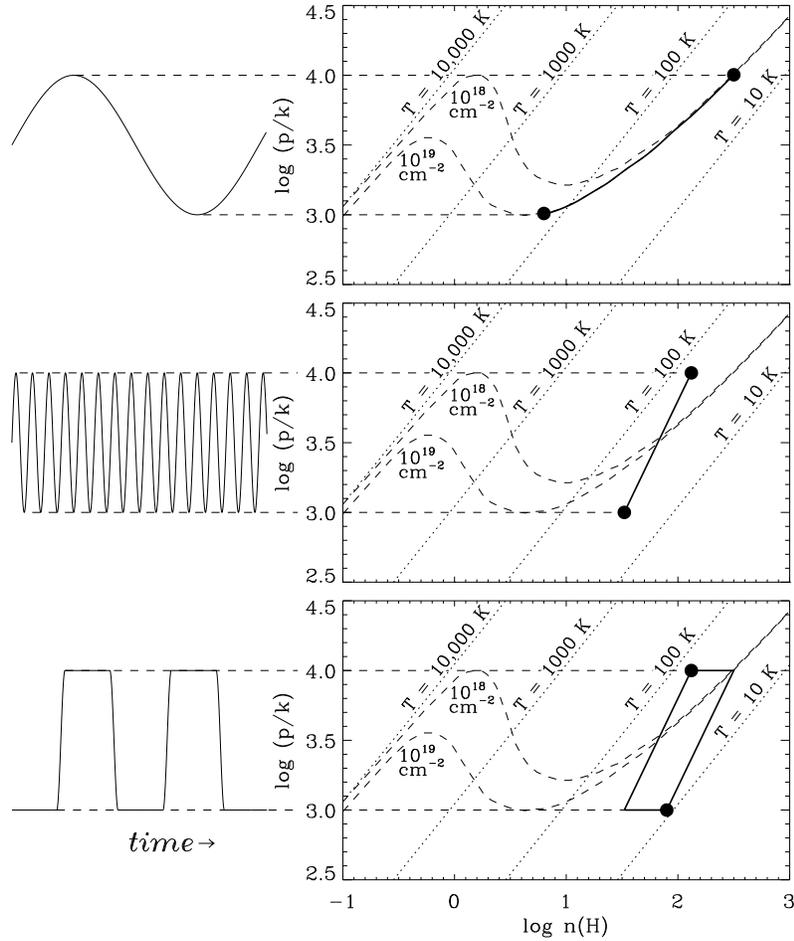,width=11cm,angle=0}}
\caption{Three schematic illustrations of how the logarithms of the ISM
pressures (ordinates) and densities (absicssae) could change for
different hypothetical driving functions of pressure as a function of
time (three periodic waves on the left).  The dashed curves inside the
panels show the thermal equilibrium curves of Wolfire et al  (1995) for
column densities $N({\rm H I})=10^{19}\,{\rm cm}^{-2}$ (lower curve) and
$10^{18}\,{\rm cm}^{-2}$ (upper curve) for shielding against external
EUV radiation.  {\it Top panel:\/} Very slow changes in pressure (shown
to the left of the panel) result in the gas conditions moving up and
down the thermal equilibrium curve with a slope in the $\log p - \log n$
plane of $\gamma_{\rm eff}\approx 0.72$.  The temperature extremes
(heavy dots) for pressures ranging from $10^3$ to $10^4\,{\rm
cm}^{-3}\,$K are 166 and 30$\,$K. {\it Middle panel:\/} Very rapid
changes in pressure will cause the gas to move on an adiabatic track
($\gamma_{\rm eff}=5/3$) with temperature extremes of 30 and 76$\,$K for
the same pressure differences. {\it Bottom panel:\/} A square-wave
driving function can achieve larger temperature extremes of 12 and
76$\,$K, as the gas is forced to move on the parallelogram track in a
clockwise fashion.}\label{examples3}
\end{figure}

Wolfire et al. (1995) have calculated thermal equilibria of the ISM
using recent refinements in the estimates for various processes that
contribute to the heating and cooling.  Two of their equilibrium curves
are shown in the three panels showing $\log (p/k)$ vs. $\log n$ of
Figure~\ref{examples3}.  If we propose that pressure changes are very
slow compared to the thermal equilibration time, then the gas will
adjust its temperature to agree with the equilibrium line as the
pressure changes from one extreme to another (solid curve).  This
condition is shown in the top panel of the figure.  However, the results
of JT01 indicate that such slow changes for the more extreme excursions
are not realistic, since large positive pressures would cause the
temperature to drop to the point that the C~I excitations become very
small, contrary to what is observed.  JT01 found that $\gamma_{\rm eff}
> 0.90$, i.e., measurably larger than $\gamma_{\rm eff}\approx 0.72$ for
the equilibrium curve.

At the opposite extreme, very rapid fluctuations in pressure will cause
the gas to move along an adiabat, where $\gamma_{\rm eff}$ becomes the
real value of $\gamma$, $c_p/c_v=5/3$, as is illustrated in the middle
panel of Fig.~\ref{examples3}.  Also, if the gas is subjected to
supersonic turbulence, shocks will form and cause momentary excursions
of the temperatures up to about 1000$\,$K for Mach numbers $M\approx
10$.

Finally, we may consider a contrived but pedagogically useful situation
where the pressure driving function is approximately a square wave. 
Here, a cycle may be established where rapid adiabatic expansions and
compressions have interludes where the gas may approach its thermal
equilibrium for the given pressure.  In this last case, somewhat larger
extremes in temperature may be achieved.  The significance of this
possiblity is discussed in the next section.

\section{The Puzzle of Cold H~I}\label{cold_HI}

Recent surveys of 21-cm emission and absorption are providing convincing
evidence that temperatures of neutral hydrogen in the ISM in many cases
deviate substantially from the values expected from thermal equilibrium 
(Gibson et al. 2000; Heiles 2001; Knee \& Brunt 2001; Dickey et al.
2002; Heiles \& Troland 2002). [The part of the equilibrium curve with a
negative slope in Fig.~\ref{examples3} is unstable to bifurcation into
separate warmer and cooler phases  (Field 1965; Shull 1987; Begelman
1990).] In a chaotic medium, one can understand why temperatures that
are intermediate between the two principal stable phases may be present. 
It is easy to imagine that mixing of warm and cool phases can operate
over timescales shorter than the time needed to segregate the gas into
stable high and low temperature branches of the equilibrium curve 
(V\'azquez-Semadeni, Gazol \& Scalo 2000; Gazol et al. 2001).  However,
H~I temperatures {\it below} a value of about 30$\,$K are more difficult
to accept.  Occasionally, temperatures as low as 10$\,$K have been
recorded.

One initially might propose that the very cold temperatures imply simply
a diminution of the heating rate or enhancement of the cooling rate. 
For instance, perhaps there is a deficit of small grains that are
responsible for photoelectric heating.  While this may sound like an
attractive solution, Wolfire et al  (1995) show that the dust-to-gas
ratio must be lower than one-tenth the normal value to have temperatures
as low as 15$\,$K, a value that is still above that of some of the 21-cm
absorbing clouds.  Many of the cold clouds seem not to be identified
with regions containing molecules which could add to the cooling rate 
(Gibson et al. 2000).

A possible solution to the cold H~I problem could arise from the
existence of mechanical cooling due to rapid pressure fluctuations.  A
single cycle of pressure changes, perhaps one that somewhat approximates
the square-wave idealization shown in the bottom panel of
Fig.~\ref{examples3}, could achieve a temperature as low as 12$\,$K soon
after a rapid expansion from a thermally stabilized condition starting
at $p/k=10^4\,{\rm cm}^{-3}\,$K.  However, in order for decompressions to drive the gas
along an adiabat, they must occur over a time scale\footnote{The
arguments presented here apply to gas that is mostly in atomic form. 
See Ballesteros-Paredes, V\'azquez-Semadeni \& Scalo  (1999) for a
similar discussion that applies to dense, molecular clouds.} that is
much shorter than the heating time,
\begin{equation}
t_{\rm heat}=(5/2){kT\over \Gamma}
\end{equation}
$\approx 5500\,{\rm yr}$ at $T=30\,$K if the heating rate
$\Gamma=6\times 10^{-26}\,{\rm erg~s}^{-1}{\rm atom}^{-1}$ that is
typical for the cold, neutral medium  (Wolfire et al. 1995).  A cloud
that is suddenly relieved of an external confining pressure will expand
only at its sound speed,
\begin{equation}
c_s=(\gamma p/\rho)^{1/2}
\end{equation}
which equals $0.54\,{\rm km~s}^{-1}$ for $T=30\,$K.  If the cloud's
shape is a sheet, then the factor of 4 drop in $n({\rm H})$ shown in
Fig.~\ref{examples3} could be accomplished in less than 5500~yr if the
thickness were less than about $10^{-3}\,$pc (cylindrical or spherical
clouds could satisfy this condition if they had diameters larger by
factors of $3^{1/2}$ and $3^{2/3}$, respectively).  However, in a medium
where there is supersonic turbulence, random inertial forces in the
fluid can create large positive mass divergences and drag a dense region
apart at an expansion rate comparable to the random velocities present,
which may exceed the sound speed by a large factor.  Thus, roughly
speaking, if the average Mach number of the turbulence is 10, we may
expect an increase of the limiting size for nearly adiabatic cooling to
$\sim$0.01~pc.

Over recent years, new findings from optical absorption lines  (Meyer \&
Lauroesch 1999; Lauroesch, Meyer \& Blades 2000; Andrews, Meyer \&
Lauroesch 2001) and 21-cm line absorption  (Dieter, Welch \& Romney
1976; Diamond et al. 1989; Frail et al. 1994; Davis, Diamond \& Goss
1996; Faison et al. 1998; Faison \& Goss 2001) have shown unmistakable
evidence that in almost any viewing direction the ISM exhibits large
contrasts in density\footnote{In interpreting raw information from
either the visible absorption line or 21-cm absorption results, one must
acknowledge that they have temperature dependences $T^{-0.62}$ and $T^{-1.00}$, respectively.} over scales ranging from $2\times 10^{-5}$ to
0.03$\,$pc.  These observations provide strong support for the pervasive
character of small structures, a conclusion that is consistent with
their being shaped by the action of turbulence.

\section{Final Remarks}
A major theme advanced in this paper is that the pressures and hence
structure of the ISM change rapidly and are probably molded by chaotic
velocity fields that converge and diverge.  This is a significant
departure from the picture that high densities arise simply from static
``clouds''  that are stabilized by a uniform external pressure.  The
notion that such clouds may be an oversimplification is not new: the
nondiscrete character of ISM phases was imagined by Chandrasekhar \&
M\"unch  (1952) when they stated, ``\dots the distribution of density is
considered to be continuous but exhibiting fluctuations of a statistical
character.''  This outlook was formed long before interstellar turbulence
was a fashionable topic.

An important consideration that arises from the dominant role of
interstellar turbulence is that under many circumstances the density
condensations are ephemeral (unless gravitational binding or the
coherent, large-scale dynamical processes listed in \S\ref{intro} become
important).  As a consequence, pressure enhancements of very short
duration may allow some of the more rapid atomic and chemical processes to
take place, but other reactions may not achieve an equilibrium state. 
For instance, the time needed to adjust the rotational temperatures of
H$_2$ range from 7000$\,$yr for moderate $J$ levels to change by 2 as
they collide with H, to about 90,000$\,$yr for transitions between ortho
and para hydrogen caused by collisions with protons.  The equilbrium
between molecular and atomic hydrogen in the general ISM has a time
constant that ranges from about 600$\,$yr for unshielded regions to over
$10^7\,$yr for shielded regions.  In view of the strong indications that
pressures can change over time scales intermediate between these
extremes, we must view with caution any interpretations that assume that
these phenomena have advanced uniformly to an equilibrium state.

\end{article}
\end{document}